\documentclass[12pt]{article}
\pdfoutput=1
\usepackage[a4paper,text={16.8cm,22.4cm}]{geometry}
\usepackage{amsmath,amsfonts,braket,slashed,amssymb,tikz,bm,psfrag,graphicx,color,dsfont,euscript}
\usepackage{multicol}
\usepackage[small,labelfont=bf]{caption}

\RequirePackage[sort&compress,square,comma,numbers]{natbib}
\allowdisplaybreaks
\newcommand{\cO}{{\mathcal O}}
\newcommand{\cC}{{\mathcal C}}
\newcommand{\cL}{{\mathcal L}}
\newcommand{\cH}{{\mathcal H}}
\newcommand{\cI}{{\mathcal I}}
\newcommand{\cA}{{\mathcal A}}
\newcommand{\cB}{{\mathcal B}}

\newcommand{\kk}{\kappa(q^2)}
\newcommand{\bb}{\beta(q^2)}

\newcommand{\be}{\begin{equation}}
\newcommand{\ee}{\end{equation}}
\newcommand{\bea}{\begin{eqnarray}}
\newcommand{\eea}{\end{eqnarray}}  
\newcommand{\no}{\nonumber}

\begin{document}

\begin{titlepage}

\begin{flushright}
\normalsize
ZH-TH~01/20\\
January 13, 2020
\end{flushright}

\vspace{1.0cm}
\begin{center}
\Large\bf\boldmath
Hunting for $B^+\to K^+ \tau^+\tau^-$ imprints  \\
on the $B^+ \to K^+  \mu^+\mu^-$ dimuon spectrum
\end{center}

\vspace{0.5cm}
\begin{center}
C.~Cornella, G.~Isidori, M.~K\"onig,  S.~Liechti, P.~Owen, N.~Serra\\
\vspace{0.7cm} 
{\em Physik-Institut, Universit\"at Z\"urich, CH-8057, Switzerland\\[3mm]}
\end{center}

\vspace{0.8cm}
\begin{abstract}
We investigate the possibility of indirectly constraining the $B^{+}\to K^{+}\tau^+\tau^-$ decay rate using precise data on the $B^{+}\to K^{+}\mu^+\mu^-$ dimuon spectrum. To this end, we estimate the distortion of the spectrum induced by the $B^{+}\to K^{+}\tau^+\tau^-\to  K^{+} \mu^+\mu^-$ re-scattering process, and propose a method to  simultaneously constrain this (non-standard) contribution and the long-distance effects associated to hadronic intermediate states. The latter are constrained using the analytic properties of the amplitude combined with data and perturbative calculations. Finally, we estimate the sensitivity expected at the LHCb experiment with present and future datasets. We find that constraints on the branching fraction of $O(10^{-3})$, competitive with current direct bounds, can be achieved with the current dataset, while bounds of  $O(10^{-4})$ could be obtained with the LHCb upgrade-II luminosity. 
\end{abstract}

\end{titlepage}

\section{Introduction}
In recent years, discrepancies between the observed values and the Standard Model (SM) predictions of the lepton-flavour universality (LFU) 
ratios $R_{D^{(\ast)}}$~\cite{Aaij:2015yra,Lees:2013uzd,Hirose:2016wfn,Aaij:2017deq,Abdesselam:2019dgh} and $R_{K^{(\ast)}}$~\cite{Aaij:2014ora,Aaij:2017vbb,Aaij:2019wad,Abdesselam:2019wac}, characterizing the semileptonic transitions $b\to c l  \nu$ and $b\to s ll$, 
have sparked great interest. The pattern of anomalies seems to  point to intriguing new-physics (NP) scenarios, with possible connections to the SM flavour puzzle.  A large class of NP models proposed to explain these hints of physics beyond the SM, and in particular those aiming for a combined explanation of the $R_{K^{(\ast)}}$ and $R_{D^{(\ast)}}$ anomalies,  imply dominant couplings to third-generation fermions, which should also enter other semileptonic $b$-quark decays.

A general expectation, confirmed by many explicit NP constructions, is that of a large enhancement of $b\to s\tau^+\tau^-$ transitions (see e.g.~\cite{  Bobeth:2011st,Glashow:2014iga, Alonso:2015sja,Barbieri:2015yvd,Crivellin:2017zlb,Buttazzo:2017ixm,Capdevila:2017iqn,Bordone:2018nbg}). While flavour-changing neutral-current (FCNC) decays with muon and electron pairs have been observed both at the exclusive and at the inclusive level, probing rare decays with a $\tau^+\tau^-$ pair in the final state is experimentally very challenging. The current experimental limits  
for all processes mediated by the $b\to s\tau^+\tau^-$ amplitude are still very far from the corresponding SM predictions~\cite{TheBaBar:2016xwe,Aaij:2017xqt}, leaving the NP expectation of possible large enhancements unchallenged.  

In this work we investigate the possibility of indirectly constraining the $b\to s\tau^+\tau^-$ amplitude via its imprint on the $B^+\to K^+ \mu^+\mu^-$ dimuon spectrum. In presence of a large NP enhancement,  the $b\to s\tau^+\tau^-$ amplitude would induce a distinctive distortion of the \mbox{$B^+\to K^+ \mu^+\mu^-$} spectrum via the (QED-induced) re-scattering process $B^+\to K^+ \tau^+\tau^- \to K^+ \mu^+\mu^-$~\cite{Bobeth:2011st}.
The latter has a discontinuity at $q^2=4m_\tau^2$ ($q^2\equiv m^2_{\mu\mu}$), namely at the threshold where the tau leptons can be produced on-shell. This gives rise to a ``cusp'' in the dimuon-invariant mass spectrum, which could in principle be detected with sufficient experimental precision. More generally, the lightness of the  $\tau$-leptons implies a well-defined deformation of the $B^+\to K^+ \mu^+\mu^-$ spectrum, which is determined only by the analytic properties of the re-scattering amplitude. 

It should be stressed that the phenomenon we are considering here is different from the QED mixing between dimension-six FCNC operators with different lepton species analysed in Ref.~\cite{Crivellin:2018yvo}. If NP is heavy and the $b\to s\tau^+\tau^-$ amplitude is strongly enhanced, the operator mixing can give rise to sizable modifications of the Wilson coefficients of the dimension-six effective Hamiltonian relevant to $b\to sl ^+l ^-$ decays ($l =e,\mu$). However, this phenomenon cannot be distinguished in a model-independent way from other NP effects of short-distance origin (at least using low-energy data only). On the contrary, the non-local effect we are interested in can be unambiguously attributed to the re-scattering of light intermediate states characterised by the tau mass, hence it can be translated into a model-independent constraint on the $B^+\to K^+ \tau^+\tau^-$ amplitude. 

The main difficulty in extracting such bound is obtaining a reliable description of the $B^+\to K^+ \mu^+\mu^-$ dimuon spectrum within the SM, or better 
in the limit where the $\tau^+\tau^- \to \mu^+\mu^-$  re-scattering is negligible. This is non trivial, given that the $B^+\to K^+ l ^+ l ^-$ spectrum 
is plagued by theoretical  uncertainties originating from $B\to K$ form factors and hadronic long-distance contributions.  While the former are smooth functions in the $q^2$ region of interest and can be well described using lattice QCD~\cite{Bouchard:2013pna,Bailey:2015dka}
 and/or light-cone sum rules~\cite{Gubernari:2018wyi}, long-distance effects induced by intermediate hadronic states, such as the charmonium resonances, are more problematic. They are genuine non-perturbative effects and introduce physical discontinuities below and above the $q^2=4m_\tau^2$ threshold. Far from the resonance region, these effects can be estimated using perturbative constraints 
derived at $q^2<0$, with $|q^2| \gg \Lambda_{\rm QCD}^2$, 
combined with a $\Lambda_{\rm QCD}^2/q^2$ or $\Lambda_{\rm QCD}^2/m_c^2$ expansion
to incorporate the leading non-perturbative corrections~\cite{Beneke:2001at,Khodjamirian:2010vf}. However, this approach is not 
suitable for our purpose, which requires a reliable description of the whole spectrum, and in particular of the resonance region. 
To achieve this goal, we adopt a data-driven approach which takes  
full advantage of the known analytic properties of the amplitude: knowing the precise location of all one- and two-particle hadronic thresholds, we use subtracted dispersion relations to describe the $q^2$--dependence of the whole spectrum in terms of a series of ($q^2$--independent) 
hadronic parameters, which are fitted from data. This method can be considered a generalisation of the approaches   
proposed in Ref.~\cite{Khodjamirian:2012rm}  and, to some extent, in Refs.~\cite{Lyon:2014hpa,Blake:2017fyh,Bobeth:2017vxj},
with a few key differences, the most notable ones being the use of subtracted dispersion relations and the explicit inclusion of two-particle thresholds.
To reduce the number of independent free parameters, perturbative constraints derived from the low-$q^2$ region are also implemented. Proceeding this way we obtain a description of the spectrum that is flexible enough to extract the 
non-perturbative parameters characterising the various hadronic thresholds from data, but retains a significant predictive power in the smooth region within 
and below the two narrow charmonium states, allowing us to set useful constraints on the 
$B^+\to K^+ \tau^+\tau^- \to K \mu^+\mu^-$ re-scattering.

The method we propose is particularly well suited for the LHCb experiment, which has already collected a large sample of $B^+\to K^+ \mu^+\mu^-$ 
events and has an excellent resolution in the dimuon spectrum~\cite{Aaij:2016cbx}.
In order to estimate the sensitivity of LHCb in view of the full run II dataset, we generate pseudo-experiments based on the yields and ampltiudes obtained in Ref.~\cite{Aaij:2016cbx}, and calculate the expected limit under the background-only hypothesis using the CLs method~\cite{CLs}.

The paper is organised as follows: in Section~\ref{sect:two} we introduce the theoretical framework necessary to describe the $B^+\to K^+ \mu^+\mu^-$ dimuon spectrum within and beyond the SM, separating short-distance contributions (Section~\ref{sect:2.1}),
long-distance contributions due to intermediate hadronic states (Section~\ref{sect:2.3}), and 
long-distance contributions due to the $\tau^+\tau^- \to \mu^+\mu^-$ re-scattering (Section~\ref{sect:tau}).
The analysis of the LHCb sensitivity is presented in Section~\ref{sect:LHCb}. The results are summarised in the Conclusions.

\section{Theoretical Framework}
\label{sect:two}

\subsection{Effective Hamiltonian and differential decay rate}
\label{sect:2.1}

The dimension-six effective Langrangian describing $b \to s ll$ transitions, renormalized at low energies [$\mu =O(m_b)$], can be decomposed as
\begin{align}
\label{eq:Leff}
 \cL_\mathrm{eff} = \frac{4G_F}{\sqrt{2}}V_{tb}V_{ts}^\ast
 \sum_{i}\cC_i(\mu)\, \cO_i\,,
\end{align}
where the leading FCNC effective operators are defined as
\begin{equation}
\begin{aligned}
\cO_7 &= \frac{e}{16\pi^2}m_b(\bar s\sigma_{\mu\nu}P_Rb)F^{\mu\nu}\,, \\ 
\cO^l_9 &= \frac{e^2}{16\pi^2}(\bar s \gamma_\mu P_L b)(\bar l \gamma^\mu l)\,, &
\cO^l_{10}&= \frac{e^2}{16\pi^2}(\bar s \gamma_\mu P_L b)(\bar l \gamma^\mu\gamma_5 l)\,,
\end{aligned}
\end{equation}
and the  most relevant four-quark operators ($q=u,c$) as
\begin{equation}
\begin{aligned}
\cO_1^q &= (\bar s \gamma_\mu P_L q)(\bar q\gamma^\mu P_L b)\,, &
\cO_2^q &= (\bar s^\alpha \gamma_\mu P_L q^\beta)(\bar q^\beta\gamma^\mu P_L b^\alpha)\,. \\
\end{aligned}
\end{equation}
Within the class of models we are considering, all relevant NP effects are encoded in the values of the Wilson coefficients $\cC^l_{7,9,10}$.  
Given the normalisation in Eq.~(\ref{eq:Leff}), $\cC^l_{7,9,10}$ and $\cC^c_{1,2}$ are real and $O(1)$ within the SM,
whereas $\cC^u_{1,2}= (V_{ub} V_{us}^*/V_{tb} V_{ts}) \times O(1)$ (see Ref.~\cite{Khodjamirian:2012rm} for the precise values of the Wilson coefficients 
and the complete basis of operators). 

The matrix elements $\langle K^+\mu^+\mu^- |  \cO_i   | B^+ \rangle$ are non-vanishing at the tree level only in the case of the FCNC operators (with $l=\mu$). 
Considering only these contributions, the  $B^+\to K^+\mu^+\mu^-$ decay rate can be written as:
\begin{eqnarray}
 \frac{d\Gamma}{dq^2} &=& \frac{\alpha_\mathrm{em}^2G_F^2 |V_{tb}V_{ts}^\ast|^2}{128\,\pi^5} \kappa(q^2) \beta(q^2) \left\{\frac{2}{3} \kappa^2(q^2) \beta^2(q^2) \left|\cC^\mu_{10}f_+(q^2)\right|^2+ \frac{m_\mu^2(m_B^2-m_K^2)^2}{q^2\, m_B^2}\left|\cC^\mu_{10}f_0(q^2)\right|^2\right.  \nonumber\\
&& \qquad\qquad\qquad  \left. + \kappa^2(q^2)\left[1-\frac{1}{3}\beta^2(q^2) \right]
 \, \left|\cC^\mu_9 f_+(q^2)+2\cC_7 \frac{m_b+m_s}{m_B
+m_K} f_T(q^2)\right|^2 \right\} \,, 
\label{eq:rate}
\end{eqnarray}
where $\kappa(q^2)=\lambda^{1/2}(m_B^2,m_K^2,q^2)/2m_B$ is the kaon momentum in the $B$-meson rest frame, $\beta(q^2)=\sqrt{1-4m_\mu^2/q^2}$, and $f_i(q^2)$ with $i=+,0,T$ are the vector, scalar and tensor $B\to K$ form factors.

\subsection{Non-local contributions: general considerations}

The non-local contributions generated by the non-leptonic operators in $ \cL_\mathrm{eff}$
and by the operator $\cO^\tau_9$ 
can be encoded in Eq.~(\ref{eq:rate}) by replacing  $\cC^\mu_9$ with a $q^2$-dependent function:
\be
 \cC^\mu_9\ \to\ \cC_9^{\mu,\mathrm{eff}}(q^2)= \cC^\mu_9+ Y_{c \bar c} (q^2) +  Y_{\rm light} (q^2) + Y_{\rm \tau \bar \tau } (q^2)\,,
 \label{eq:C9eff}
\ee
where $Y_{\cI} (q^2)$ denotes the non-local contributions corresponding to the intermediate state $\cI$, which can annihilate into a dimuon pair via a single-photon exchange.

The functions $Y_{c \bar c} (q^2)$ and  $Y_{\rm light} (q^2)$ encode non-perturbative hadronic contributions, which cannot be estimated reliably in perturbation theory, at least in a large fraction of the accessible $q^2$ spectrum. 
Adopting a notation similar to that of Ref.~\cite{Khodjamirian:2012rm}, we can express $Y_{c \bar c} (q^2)$ as
\be
Y_{c \bar c} (q^2) =  \frac{16\pi^2  }{ f_+(q^2) }   \cH_{c \bar c}^{(BK)}(q^2), 
\label{eq:Ydef}
\ee
where $\cH_{c \bar c}^{(BK)}(q^2)$ is defined by 
the gauge-invariant decomposition of the following non-local hadronic matrix element
\bea
i \int d^4 x e^{iq\cdot x} \langle K(p) | T\left\{ j_\mu^{\rm em}(x),  \sum_{i=1,2}   \cC^c_i \cO^c_i   \right\} | B(p+q) \rangle
= \left[ (p\cdot q) q_\mu - q^2 p_\mu \right] \cH_{c \bar c}^{(BK)}(q^2)~,
\label{eq:HBKdef}
\eea
with $j_\mu^{\rm em}= \sum_{q} Q_q \bar q \gamma^\mu q$.
The function $Y_{\rm light} (q^2)$, containing the contribution of the subleading non-leptonic operators in $ \cL_\mathrm{eff}$, is defined in a similar way via the replacement 
\be
 \sum_{i=1,2}   \cC^c_i \cO^c_i    ~\to  \sum_{i=3-6,8}   \cC_i \cO_i  + \sum_{i=1,2}   \cC^u_i \cO^u_i~.
 \label{eq:Olight}
\ee

\begin{figure}
 \centering
 \raisebox{0ex}{\includegraphics[scale=.6]{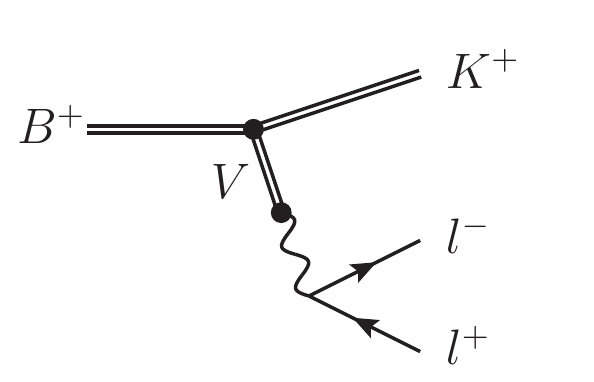}}\hspace{8ex} \includegraphics[scale=.6]{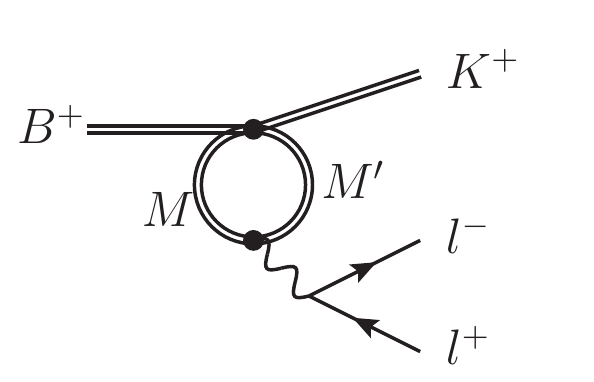}
 \caption{Diagrammatic representations of the long-distance contributions to $\cC_9^\mathrm{eff}$. The left-hand side depicts the exchange of a single vector resonance. The graph on the right-hand side shows the contribution from two-particle intermediate states.\label{fig:longdist}}
\end{figure}

Our main strategy is to write the non-perturbative functions  $Y_{c \bar c} (q^2)$ and  $Y_{\rm light} (q^2)$
using hadronic dispersion relations.  More precisely, for the leading 
charm contribution we consider one- (1P) and two-particle (2P) intermediate states (see Fig.~\ref{fig:longdist}),  using dispersion relations subtracted at $q^2=0$, while for the subleading $Y_{\rm light} (q^2)$ function we consider only  one-particle intermediate states and use unsubtracted dispersion relations.  \\
Under these conditions,  
$\cC_9^{\mu,\mathrm{eff}}(q^2)$ in Eq.~(\ref{eq:C9eff}) is finally decomposed according to
\be
\cC_9^{\mu,\mathrm{eff}}(q^2)=\cC^\mu_9 +  Y^{(0)}_{c\bar c} + \Delta Y^{\rm 1P}_{c \bar c} (q^2) 
+ \Delta Y^{\rm 2P}_{c \bar c} (q^2) + Y^{\rm 1P}_{\rm light} (q^2)  + Y_{\rm \tau\bar\tau} (q^2)~,
 \label{eq:C9eff2}
\ee
with $\Delta Y^{\rm 1P}_{c \bar c} (0) = \Delta Y^{\rm 2P}_{c \bar c} (0) = 0$.  

In the next section we analyse the structure of  $\Delta Y^{\rm 1P}_{c \bar c} (q^2)$, $\Delta Y^{\rm 2P}_{c \bar c} (q^2)$,
and $Y^{\rm 1P}_{\rm light} (q^2)$ in detail. The expression of  $Y_{\rm \tau\bar\tau} (q^2)$, which is the only term in 
Eq.~(\ref{eq:C9eff2}) that can be fully evaluated in perturbation theory, is given in Sect.~\ref{sect:tau}.

\subsection{Long-distance hadronic contributions}
\label{sect:2.3}

The general structure of the subtracted dispersion relation used to determine $\Delta Y_{c \bar c} (q^2)$ is
\bea
\Delta Y_{c \bar c} (q^2) =  \frac{16 \pi q^2}{  f_+(s)}  \int_{m^2_{J/\Psi}}^\infty  ds \frac{ 1  }{s (s - q^2)  } ~ \frac{1}{2i}   {\rm Disc}\left[  \cH_{c \bar c}^{(BK)}(s)    \right] 
\equiv  \frac{q^2}{\pi}  \int_{m^2_{J/\Psi}}^\infty  ds   \frac{ \rho_{c\bar c}(s)  }{s (s - q^2)  }\,.
\eea
The function $\rho_{c\bar c}(s)$ is the spectral-density function describing the hadronic states $\cI_{c\bar c}$, 
characterized by valence charm-quarks and invariant mass $s$, 
contributing as real intermediate states in the re-scattering $B\to K \cI_{c\bar c} \to K \mu^+\mu^-$.
As noted before, we decompose $\rho_{c\bar c}(s)$ into one- and two-particle intermediates states,
$\rho_{c\bar c}(s)  = \rho^{\rm 1P}_{c\bar c}(s)  + \rho^{\rm 2P}_{c\bar c}(s)$,  
\bea 
 \rho^{\rm 1P}_{c\bar c}(s) &\propto&  \sum_j  \cA( B\to K V^0_j) \delta( s - m_j)~,   \label{eq:rho1P}  \\
\rho^{\rm 2P}_{c\bar c}(s) &\propto&  \sum_j  \int d p_j^2 ~\delta( s- p_j^2 )  \int \frac{d^3 \vec{p}_{j_1} d^3 \vec{p}_{j_2}   }{ 16 \pi^2  E_{j_1}  E_{j_2}  }  
    \cA( B\to K M^+_{j_1} M^-_{j_2} )  \delta^{(4)}( p_j - p_{j_1} -p_{j_2}) ~,
\eea
neglecting the phase-space suppressed contribution with three or more particles. 

\subsubsection{Charmonium resonances}

For the sake of simplicity, in  Eq.~(\ref{eq:rho1P}) we have treated the single-particle states as infinitely narrow resonances. The effect of finite widths can be 
incorporated via Breit-Wigner functions, yielding 
\be
\Delta Y_{c\bar c}^\mathrm{1P}(q^2) =  \sum_{j = \Psi(1S),  \ldots,\Psi(4415) }  \eta_j\, e^{i\delta_j} \frac{q^2}{m_j^2} 
A^\mathrm{res} _j (q^2)~,
\qquad 
A^\mathrm{res} _j (s) = \frac{m_j\Gamma_j}{(m_j^2-s)-im_j \Gamma_j}\,~,
\label{eq:1P}
\ee
where the sum runs over all the charmonium vector resonances in the accessible kinematical range. 
Here $\eta_j$ and $\delta_j$ are real parameters which must be determined from data, similarly to 
what has been performed by the LHCb collaboration in~\cite{Aaij:2016cbx}.
The fitted $\eta_j$'s can be put in one-to-one correspondence with the product of 
the $B^+ \to K^+ V^0_j$  and $\ V^0_j \to \mu^+\mu^- $ branching fractions via 
\bea
&& \cB( B^+ \to K^+ V^0_j) \times \cB(  V^0_j \to \mu^+\mu^-) = \tau_{B^+} \frac{G_F^2\alpha^2|V_{tb}V_{ts}^*|^2}{128\pi^5}  
  \int\limits_{4 m^2_\mu}^{(m_B - m_K)^{2}}  dq^2 
        \kk^3 \times \no \\  
 && \qquad\qquad    \times  \left[ \bb - \frac{1}{3}\bb^3 \right]     \left| f_+(q^2) \right|^2 \left| \eta_j \right|^2 \left| \frac{q^2}{m_j^2}  A_j^{\rm res} (q^2) \right|^2  \,.
    \label{eq:br:constraint}
\eea

The expression (\ref{eq:1P}) differs from the decomposition adopted in Ref.~\cite{Aaij:2016cbx} by the  $q^2/m^2_j$ term, 
which arises from the subtraction procedure in the dispersion relation.
On the one hand, the use of subtracted dispersion relations for the charm contribution is necessary to ensure the convergence 
of the integral in the two-particle intermediate states (see sect.~\ref{sect:2body}).  On the other hand, 
choosing the subtraction point at $q^2=0$ allows us to decouple 
the determination of the resonance parameters of 
the spectrum from the overall normalisation of the rate, and hence from the determination of $\cC_9^\mu$ from data.
The price to pay is the appearance of the undetermined constant term $Y^{(0)}_{c\bar c}= Y_{c \bar c} (0)$ in Eq.~(\ref{eq:C9eff2}).
This term plays no role  in the description of the dimuon spectrum, but is relevant for the extraction of the value of $\cC_9^\mu$.
To this purpose, we note that the estimate presented in Ref.~\cite{Khodjamirian:2012rm},
which is based on a $\Lambda^2/m_c^2$ expansion and also takes next-to-leading $O(\alpha_s)$ corrections 
on the pure partonic result into account (see sect.~\ref{sect:pert}), yields
\be
Y^{(0)}_{c\bar c} \approx -0.10 \pm 0.05~,
\ee 
which is about $-(2\pm1)\%$ of $\cC_9^{\mu, {\rm SM}} \approx 4.23$. 

\subsubsection{Two-particle intermediate states}
\label{sect:2body}

Proceeding in a similar way, we can decompose the two-particle contributions as 
\begin{align}
 \Delta Y^\mathrm{2P}_{c\bar c}(q^2)
 =\sum_{j} \eta_j e^{i\delta_j} A_j^\mathrm{2P}(q^2)\,, \qquad A_j^\mathrm{2P}(q^2) = \frac{q^2}{\pi}\int_{s_0^j}^\infty \frac{d s}{s}\frac{
 \hat \rho_j (s)}{(s-q^2)}\,,
 \label{eq:charm2p}
\end{align}
where $\hat \rho_j(s)$ are normalised spectral densities for the two-body intermediate states characterised by the threshold $s_0^j = (m_{j_1}+m_{j_2} )^2$. 

\begin{figure}
\centering
 \includegraphics[scale=.6]{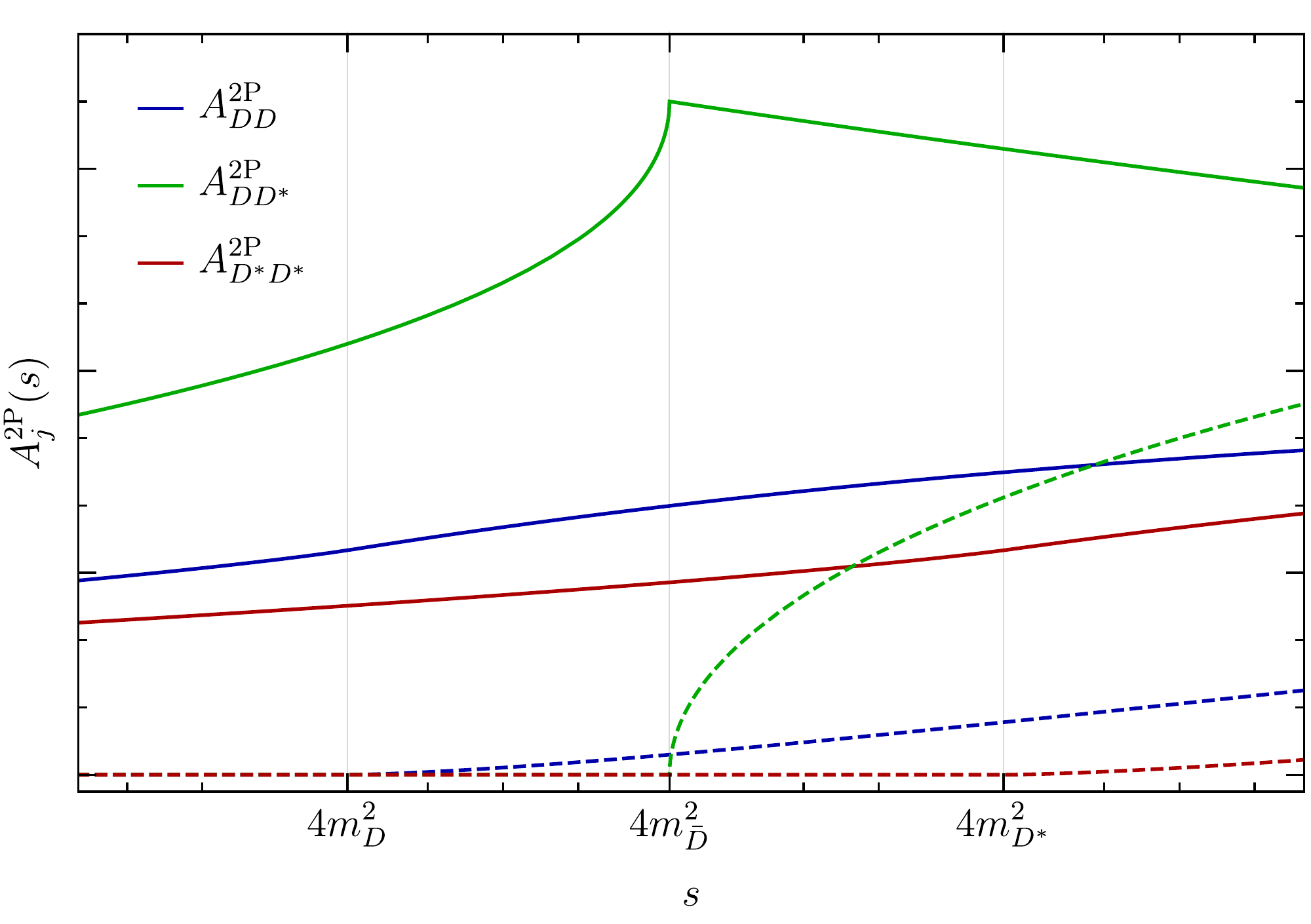}
 \caption{Real (solid) and imaginary (dashed) parts of the normalised hadronic two-particle contributions to 
 $Y_{c\bar c}(q^2)$,  as defined in Eq.~(\ref{eq:charm2p}).}
 \label{fig:charm2p}
\end{figure}

While we do not have a precise estimate of these spectral densities at generic kinematical points, an excellent description of their behaviour around the respective thresholds is obtained by approximating them with powers of  the K\"all\'en function, 
with an exponent determined by the lowest  partial wave allowed in the $B^+ \to K^+  M_1 M_2  \to K^+ \mu^+\mu^-$ re-scattering. 
This is  because higher-order partial waves, characterised by higher powers of the K\"all\'en function,
are both phase-space suppressed and, most importantly, give rise to a less singular 
behaviour at the threshold. From angular momentum conservation we can then determine the leading partial wave and obtain the following estimates for the normalised spectral densities of the two-particle intermediate states of lowest mass:
\begin{align}
 \hat \rho_{DD}(s) &= \left( 1-\frac{4m_D^2}{s}\right) ^{3/2}\,, &
 \hat \rho_{D^\ast D^\ast}(s) &= \left( 1-\frac{4m_{D^\ast}^2}{s}\right) ^{3/2}\,, &
\hat  \rho_{DD^\ast}(s) &= \left( 1-\frac{4m_{\bar D}^2}{s}\right) ^{1/2}\,.
\end{align}
In the case of the $DD^\ast$ intermediate state we have replaced the complete expression depending on both masses with 
a simplified one depending only on $m_{\bar D}= (m_D+m_{D^\ast})/2$, which provides an excellent approximation.  
With these estimates in place we find:
\begin{align}
\Delta Y^{2\mathrm{P}}_{c\bar c} (q^2)  =    \eta_{\bar D} e^{i\delta_{\bar D}}  h_{S} \left(m_{\bar D},  q^2 \right) + 
\sum_{j=D, D^*}  \eta_{j} e^{i\delta_j}  h_{P} \left( m_{j},  q^2 \right)\,, 
\label{eq:2P}
\end{align}
with
\begin{equation}
\begin{aligned}
h_P \left(m,  q^2 \right) &= \frac{2}{3}+ \left( 1 -\frac{4 m^2}{q^2} \right) h_S \left(m,  q^2 \right) \,, \qquad 
h_S \left(m,  q^2 \right) &= 2-  G\left( 1 -\frac{4 m^2}{q^2} \right) \,,  \label{eq:loopF}
\end{aligned}
\end{equation}
and 
\begin{align}
 G(y) =  \sqrt{|y|} \left\{ \Theta(y) \left[ \ln\left( \frac{1+\sqrt{y}}{ 1-\sqrt{y} } \right)-i \pi \right]  
+ 2~\Theta(-y) \arctan\left( \frac{1}{\sqrt{-y}} \right) \right\}\,.  
\end{align}
It is worth noting that, while the lowest threshold is at $q^2=4m_D^2$, the contribution from the $DD^\ast$ intermediate state is 
the only one which can occur in the $S$-wave, corresponding to a singular (square-root) behaviour at the threshold (see Fig.~\ref{fig:charm2p}).

\subsubsection{Light resonances}
\label{sect:light}
The remaining hadronic contribution we need to estimate is $Y_{\rm light} (q^2)$,  defined by 
Eqs.~(\ref{eq:Ydef})--(\ref{eq:HBKdef}) via the replacement (\ref{eq:Olight}).
The Wilson coefficients of the effective operators appearing in $\cH_{\rm light}^{(BK)}(q^2)$ are either loop- or CKM-suppressed.
As a result, we can limit ourselves to include only one-particle 
hadronic intermediate states. In principle, such operators describe transitions also 
to states with valence charm quarks; however, since we fit the
hadronic  coefficients $\eta_j$ from data, these terms are naturally 
absorbed in the  $\eta_j$  appearing in  $\Delta Y_{c\bar c} (q^2)$. 
We are thus left only with vector resonances containing 
light valence quarks. Among them, we can further restrict 
the attention to the $\rho$, $\omega$,  and $\phi$ resonances, since the leptonic decay rates of the heavier states are very small.

There is no clear advantage in using subtracted vs.~unsubtracted dispersion relations
in describing the contributions of the light vector resonances. The convergence of the dispersive integrals
does not pose a problem, and the subtraction at  $q^2=0$ is not particularly useful since
the light-quark contributions are in a non-perturbative regime at $q^2=0$. However, when fitting data, 
the subtraction at  $q^2=0$ retains the advantage of decoupling the determination 
of the spectrum from that of the Wilson coefficient. As default option, we adopt unsubtracted dispersion relations. As discussed in Sect.~\ref{sect:model}, checking the stability of the result using subtracted vs.~unsubtracted dispersion relations for the light vector resonances provides an estimate of the ``model error" of the proposed approach.

Given these considerations, we decompose $Y^{\rm 1P}_{\rm light} (q^2)$ as
\be
Y_{\rm light}^\mathrm{1P}(q^2) =  \sum_{j = \rho, \omega, \phi }  \eta_j\, e^{i\delta_j} 
A^\mathrm{res} _j (q^2)~,
\label{eq:1P_light}
\ee
in perfect analogy with the decomposition adopted in Ref.~\cite{Aaij:2016cbx} 
for these light states.

 \subsubsection{Theoretical constraints on the hadronic parameters}
 \label{sect:pert}

The hadronic decompositions in Eqs.~(\ref{eq:1P}), (\ref{eq:2P}) and~(\ref{eq:1P_light})
contain 12 free complex parameters: 6 
in $\Delta Y_{c\bar c}^\mathrm{1P}(q^2)$, 3  in $\Delta Y_{c\bar c}^\mathrm{2P}(q^2)$,
and 3  in $Y_{\rm light}^\mathrm{1P}(q^2)$. In principle, since they correspond 
to different functional forms, they could all be fitted from data. In practice however, an unconstrained fit would leave 
significant degeneracies in the parameter space. It is therefore useful to restrict the 
variability of such parameters using theoretical constraints.
In the following we discuss three conservative conditions which can be imposed 
using perturbative arguments. 

\paragraph{I.} {\em Constraint on the slope of $\Delta Y_{c \bar c} (q^2)$ at $q^2=0$.}\\
The lowest-order perturbative estimate of $\Delta Y_{c \bar c} (q^2)$ is 
obtained by factorising the matrix element $\langle K(p) | \bar s \gamma^\mu b | B(p+q) \rangle$ 
in Eq.~(\ref{eq:HBKdef}) and computing the charm-loop 
at $O(\alpha^0_s)$:
\bea
\Delta Y^{\rm pert}_{c \bar c} (q^2) &=& 2  \left(\cC_2+\frac{1}{3} \cC_1\right)  \times Q_c \times q^2 \int_{4 m_c^2}^\infty  ds
\frac{\sqrt{1-\frac{4 m_c^2}{s} }\left(1+\frac{2 m_c^2}{s}\right) }{s (s - q^2)  } \no \\
&=&  2 \left(\cC_2+\frac{1}{3} \cC_1\right)  \left[ h_S(m_c,q^2)-\frac{1}{3}h_P(m_c,q^2) \right] \,.
\eea
This expression is certainly not a good approximation of $\Delta Y_{c \bar c} (q^2)$ close to the resonance region;
however, it is expected to provide a reasonable approximation at $q^2 \approx  0$,
up to $O(\Lambda_{\rm QCD}/m_c^2)$ corrections. We can thus use it to set bounds on the slope of 
$\Delta Y_{c \bar c} (q^2)$  in the vicinity of $q^2=0$. The perturbative result implies
\be
\left. \frac{d}{ dq^2} \Delta Y^{\rm pert}_{c \bar c} (q^2) \right|_{q^2=0} = 
\frac{4}{15}\left(\cC_2 + \frac{1}{3}\cC_1 \right)  \frac{1}{m_c^2} \approx (1.7\pm 1.7) \times 10^{-2} ~{\rm GeV}^{-2}~,
\label{eq:Pslope}
\ee
where the numerical value has been obtained setting $m_b/2 < \mu < 2m_b$ and $m_c=1.3$~GeV.
According to the analysis of  Ref.~\cite{Khodjamirian:2012rm}, the inclusion of  $O(\Lambda_{\rm QCD}/m_c^2, \alpha_s)$ 
corrections (which involve new hadronic matrix elements) modifies the above prediction to $-(0.5 \pm 0.2) \times 10^{-2}~{\rm GeV}^{-2}$. 
Given these considerations, in the numerical analysis we employ the following constraints
\bea
{\rm Re}\left[ \sum_{j=\Psi(1S),\ldots} \eta_{j} e^{i \delta_j} \frac{\Gamma_j}{m^3_j}  +
 \eta_{\bar D} e^{i \delta_j}   \frac{1}{6 m_{\bar D}^2} +
\sum_{j=D, D^*}  \eta_{j}  e^{i \delta_j}   \frac{1}{10 m_j^2}  \right]  &=& (1.7\pm 2.2) \times 10^{-2}~{\rm GeV}^{-2}~, \no \\
\left| \sum_{j=\Psi(1S),\ldots} \eta_{j} e^{i \delta_j} \frac{\Gamma_j}{m^3_j}  +
 \eta_{\bar D} e^{i \delta_j}   \frac{1}{6 m_{\bar D}^2} +
\sum_{j=D, D^*}  \eta_{j}  e^{i \delta_j}   \frac{1}{10 m_j^2}  \right|  &\leq&  5 \times 10^{-2}~{\rm GeV}^{-2}~,
\label{eq:THbound1}
\eea
where we slightly enlarged the error from (\ref{eq:Pslope}), such that the $1\sigma$ range covers the 
difference between the central value in  (\ref{eq:Pslope}) and the one 
including  $O(\Lambda_{\rm QCD}/m_c^2, \alpha_s)$ corrections estimated in Ref.~\cite{Khodjamirian:2012rm}.

\paragraph{II.} {\em Upper bound on the $|\eta_j|$ in $\Delta Y^{\rm 2P}_{c \bar c} (q^2)$.}\\
The comparison of the perturbative result with $\Delta Y^{\rm 2P}_{c \bar c} (q^2)$ also
allows us to define the natural range for the $\eta_{\bar D, D, D^*}$ parameters,
which are poorly constrained by data. Focusing the attention on the leading 
$S$-wave contribution, it turns out that the 
perturbative quark loop can be saturated by the $DD^*$ meson loop, 
in the limit $m_c \to  m_{\bar D}$, setting $\eta_{\bar D} = 2 (\cC_2 + \cC_1/3 ) \approx (0.2 \pm 0.2).$
On general grounds, each of the exclusive meson contributions should be significantly smaller than the inclusive quark contribution. As a result, in the following we set the 
upper limit
\be
\left| \eta_{\bar D, D, D^*} \right| \leq 0.2~.
\label{eq:THbound2}
\ee

\paragraph{III.} {\em Upper bound on $|Y^{\rm 1P}_{\rm light} (q^2=0)|$.}\\
Using an unsubtracted dispersion relation and taking into account only one-particle intermediate states for the light-quark contributions implies  $Y^{\rm 1P}_{\rm light} (q^2) \to 0 $ for large $q^2$, while $Y^{\rm 1P}_{\rm light} (0) \not= 0$. More precisely, one finds a power-like suppression of the type $Y^{\rm 1P}_{\rm light} (q^2) \sim  \Lambda_{QCD}^2/q^2$ at large $q^2$,
whereas $Y^{\rm 1P}_{\rm light} (0)$ is  not parametrically suppressed by any scale ratio. 
However, since  the Wilson coefficients entering 
$Y^{\rm 1P}_{\rm light}$ are  strongly suppressed,  either by loop factors or by subleading CKM factors,
$|Y^{\rm 1P}_{\rm light} (0)|$  cannot be too large. 
Parametrically we expect
\be
| Y^{\rm 1P}_{\rm light} (0) | <   O(1) \times  {\rm max}  \{ |\cC_{3\ldots 6}|,  | \cC_{1,2}^{u} | \}~.
\ee
Taking the size of the $\cC_i$ into account, we set the conservative bound\footnote{
The largest perturbative contribution is the one induced by strange-quark loops, yielding 
$\Delta Y^{\rm pert}_{s \bar s} (q^2) = \cC_s  [h_S(m_s,q^2)-\frac{1}{3}h_P(m_s,q^2)]$,
with $|\cC_s| = (2/3) | 4 \cC_3 +4  \cC_4 + 3 \cC_5 + \cC_6 |  \approx 0.05 \pm 0.02$. }
\be
\left| Y^{\rm 1P}_{\rm light} (0) \right|     \approx 
 \left|  \sum_{j = \rho, \omega, \phi }  \eta_j  \frac{\Gamma_j}{m_j}    \right| \leq 0.1~,
 \label{eq:THbound3}
\ee
which should be interpreted as a constraint on the relative phases of the light resonances.

 \subsubsection{Estimate of the ``model error"}
 \label{sect:model}
Despite not being entirely dictated by first principles, the parameterisation of
long-distance hadronic contributions discussed so far contains all the 
relevant one- and two-particle discontinuities of the amplitude, 
with free coefficients to be fixed by data. It should therefore 
provide a sufficiently general (and unbiased) description of the impact of hadronic 
contributions on the $B^+\to K^+ \mu^+\mu^-$ spectrum. Still, it may be worthwhile to assess whether the proposed parameterisation influences the extraction of information on $Y_{\rm \tau\bar\tau} (q^2)$ and, correspondingly, the extraction of a bound on $\cB(B^+\to K^+ \tau^+\tau^-)$. 
An estimate of this ``model error" can be obtained by examining the stability of the obtained bound on $\cB(B^+\to K^+ \tau^+\tau^-)$ under small variations of the model assumptions.
The latter include: i)~the use of subtracted vs.~unsubtracted dispersion relations for the light resonances; 
ii)~the use of $q^2$-dependent widths for both charmonium and/or light resonances; iii) strengthening or
relaxing the theoretical constraints in Eqs.~(\ref{eq:THbound1}), (\ref{eq:THbound2}), and (\ref{eq:THbound3}).

\subsection{Tau-lepton contribution}
\label{sect:tau}

The contribution from the intermediate $\tau$-leptons can be computed in perturbation theory, yielding 
\begin{align}
 Y_{\tau\bar\tau}(q^2)=-\frac{\alpha_\mathrm{em}}{2\pi}\,\cC_9^\tau \,
 \left[ h_S(m_\tau,q^2)-\frac{1}{3}h_P(m_\tau,q^2) \right]\,,
\end{align}
with the functions $h_L(m,s)$ defined in Eq.~\eqref{eq:loopF}. 
The functional form is identical to the one of the perturbative charm contribution and,
to a large extent, to the one of the $DD^*$ contribution, illustrated in Fig.~(\ref{fig:charm2p}).
However, the cusp is located at $q^2=4m_\tau^2$, sufficiently well separated from the
various hadronic thresholds. 

In principle, the short-distance $b\to s\tau^+\tau^-$ amplitude does not need to be controlled by the CKM matrix in a generic NP model. However, in most realistic scenarios the weak phases of all $b\to s l ^+ l ^-$ amplitudes are aligned to the SM one, implying ${\rm Im}(\cC_9^\tau)= {\rm Im}(\cC_9^\mu)=0$. In the following, we adopt this (motivated) simplifying assumption.

An estimate of the maximal allowed size of $|\cC_9^\tau|$ can be derived
from the experimental upper bound on $\cB(B^+ \to K^+ \tau^+\tau^-)< 2.25\times 10^{-3}$ at 90\% CL by Babar~\cite{TheBaBar:2016xwe},
which is more than four orders of magnitude larger than $\cB(B^+ \to K^+ \tau^+\tau^-)_\mathrm{SM}\approx 1.5 \times 10^{-7}$~\cite{Bouchard:2013mia}.  
Neglecting the contributions from operators other than $\cO^\tau_9$ and $\cO^\tau_{10}$, we find 
\begin{align}
\cB(B^+ \to K^+ \tau^+\tau^-)  \approx 
 \left\{  \begin{array}{ll}  
 8.7 \times 10^{-9}  \times | \cC^\tau_{9} |^2   \qquad   &  \cC^\tau_9=\cC^\tau_{10}~,\\
2.7 \times 10^{-9}  \times | \cC^\tau_{9} |^2 \qquad   &  \cC^\tau_{10}=0~. \\
\end{array}  \right. 
\label{eq:C9taumax}
\end{align}
In the case $\cC^\tau_9=\cC^\tau_{10}$ ($\cC^\tau_{10}=0$) the Babar result then implies $ |\cC^\tau_{9} |  \leq 5.1 \times 10^2~
(9.1 \times 10^2)$, to be compared to $\cC_9^{\tau,\mathrm{SM}}\approx 4.2$.
As we discuss below, saturating this bound leads to a pronounced ditau cusp in the spectrum (see Figure~\ref{fig:pseudodata}), opening the possibility of extracting a more stringent 
bound  on $\cB(B^+ \to K^+ \tau^+\tau^-)$ from a precise measurement of the $B^+\to K^+ \mu^+\mu^-$  dimuon spectrum.

\section{Analysis of the expected sensitivity at LHCb}
\label{sect:LHCb}

\begin{figure}
\centering
 \includegraphics[width=0.8\textwidth]{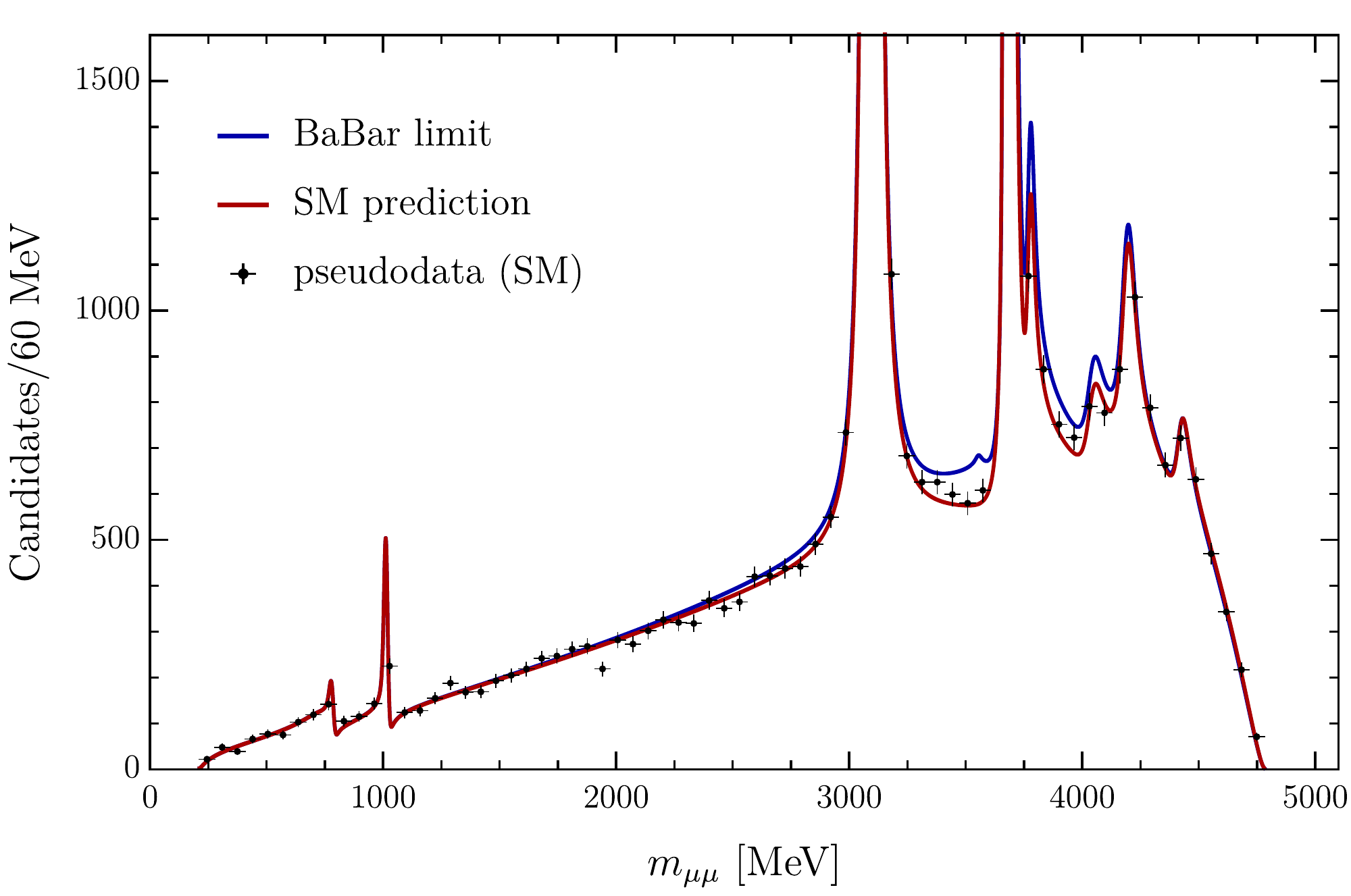}
 \caption{Example pseudodata expected from the full run II dataset collected by the LHCb experiment assuming the SM. 
The distribution expected if the $B^+\to K^+ \tau^+\tau^-$ branching fraction were present at the current experimental limit of $2.25\times 10^{-3}$ is overlaid.}
 \label{fig:pseudodata}
\end{figure}

In order to assess the sensitivity to the branching ratio $\cB(B^+ \to K^+ \tau^+\tau^-)$ at the LHCb experiment, we generate pseudo-experiments 
corresponding to the signal yields obtained in Ref.~\cite{Aaij:2016cbx} and scaled to the full run II dataset, 
taking into account the collected luminosity and \mbox{$b$-hadron} cross-section increase at 13 TeV~\cite{Aaij:2016avz}. 
This leads to around 40,000 non-resonant \mbox{$B^{+}\to K^{+} \mu^{+}\mu^{-}$} candidates. As the efficiency is reasonably flat as a 
function of dimuon mass and the background level is very low, we neglect these effects. Fig.~\ref{fig:pseudodata} shows the fit model with a dataset generated at the expected yield. This illustrates the visible sensitivity to a hypothetical 
signal component generated according to the current experimental limit~\cite{TheBaBar:2016xwe}.

The size and phase of the one-particle resonant contributions are determined from the branching fractions reported in Ref.~\cite{Aaij:2016avz},
which are used to determine the initial values of $\eta_j$ and $\delta_j$ for the data to be generated. Due to the complicated experimental resolution
effects near the $J/\psi$ and $\psi(2S)$ resonances, the regions $9.2 < q^{2} < 10.0\,{\rm GeV}^{2}/c^{4}$ and $13.2 < q^{2} < 13.95 \,{\rm GeV}^{2}/c^{4}$
are excluded from the fit and the phase differences associated with these 
resonances are constrained to the uncertainties in Ref.~\cite{Aaij:2016cbx}. Outside of this region, finite-resolution effects in $q^2$ are ignored
as all the components are broad. In order to mimic the sensitivity one would have when fitting the data, Gaussian constraints are applied to the $J/\psi$ and $\psi(2S)$ resonant parameters according to the uncertainties reported in Ref.~\cite{Aaij:2016avz}.

The component which most closely resembles the signal is the contribution from two-particle hadronic intermediate states. We allow the phases and magnitudes of these states to vary in the fit. As the shape of the $\hat{\rho}_{DD}$ and $\hat{\rho}_{D*D*}$ spectral densities are very similar, we combine them with an equal contribution to avoid large correlations in the fit. The correlation coefficient between the two-particle contribution and the signal is around $0.6$, reflecting their similar shapes. 

The form factor uncertainties are taken from Ref.~\cite{Gubernari:2018wyi} 
 and are implemented in the fit as a multivariate Gaussian constraint. The data slightly helps constrain the form factor parameters, but this affects the sensitivity on $\cC_9^\tau$ only in a mild way.

As discussed above, a possible $B^+\to K^+ \tau^+\tau^- \to K^+ \mu^+\mu^-$ re-scattering leads to two features in the $B^+\to K^+\mu^+\mu^-$ dimuon spectrum. One is the cusp in-between the $J/\psi$ and $\psi(2S)$ resonances, the other is a distortion in the shape of the non-resonant $B^{+}\to K^{+} \mu^{+}\mu^{-}$ component. The cusp is not the most sensitive signature of the ditau signal, due to its relatively small contribution. Instead, it is the shape of the non-resonant part which generates the largest sensitivity. Consequently, neglecting the resolution is justified and the shape of the charmonium contribution is important in constraining a possible $B^+\to K^+ \tau^+\tau^-$ signal.

The expected sensitivity on the $\cC_9^\tau$ contribution is determined using the $CL_{s}$ method~\cite{CLs}. The sensitivity with the current dataset is reported in Table~\ref{tab:sensitivity}, along with two other potential 
future scenarios corresponding to the LHCb upgrade-II luminosity and a hypothetical improvement of the form factor uncertainties by a factor of three. The estimated sensitivity utilising the run I-II datset corresponds to a limit on the $B^+\to K^+\tau^+\tau^-$ branching ratio which is slightly more stringent than the current constraints placed by the BaBar collaboration and is expected to compete with the projected sensitivity of the Belle-II experiment when more data is collected.

  \begin{table}[tb]
\centering
  \resizebox{\textwidth}{!}{  
    \begin{tabular}{ c  c  c  c}
      \hline
      Scenario & $\cC_9^\tau$ (90\% CL) & $\mathcal{B}$  ($C_{10\tau}=-C_{9\tau}$) & $\mathcal{B}$  ($C_{10\tau}=0$)    \\
      \hline
       Run I-II dataset & 533 & $2.7\times 10^{-3}$& $0.8\times 10^{-3}$ \\ 
       Run I-V dataset & 139 & $1.8\times 10^{-4}$& $0.5\times 10^{-4}$ \\ 
       Run I-II dataset, improved form factors & 533 & $2.7\times 10^{-3}$& $0.8\times 10^{-3}$ \\ 
       Run I-V dataset, improved form factors & 127 & $1.5\times 10^{-4}$& $0.5\times 10^{-4}$ \\ 
      \hline
    \end{tabular}}
\caption{Sensitivity to $C_{\tau\tau}$ according to various LHCb scenarios.}
\label{tab:sensitivity}
\end{table}

\section{Conclusions}

If the branching ratio $\mathcal{B}(B^+ \to K^+\tau^+\tau^-)$ were significantly enhanced over its SM value, it would induce a peculiar distortion of the $B^+\to K^+ \mu^+\mu^-$ spectrum, characterised by a cusp at $q^2=4m_\tau^2$ and by a distortion of the dimuon distribution.
In this work we have proposed a method that uses this effect as a tool to extract 
a bound on $\cB(B^+ \to K^+\tau^+\tau^-)$ from future precise measurements of 
 ${\rm d}  \Gamma (B^+\to K^+ \mu^+\mu^-)/ {\rm d} q^2$.

A necessary ingredient to achieve this goal is a reliable description of the $B^+\to K^+ \mu^+\mu^-$ dimuon spectrum, within the SM,
in the full kinematical range, especially in the region of the narrow charmonium states. 
As we have shown, this can be obtained by means of a data-driven approach which takes full advantage of the known analytic properties of the decay 
amplitude, supplemented by robust theoretical constraints. 
Our approach differs from previous attempts of including non-local hadronic 
contributions to the $B^+\to K^+ \mu^+\mu^-$  decay amplitude by three main points: i) the use of dispersion relations 
subtracted at $q^2=0$ for the charmonium states; ii) the inclusion of two-particle thresholds; iii) the use of short-distance constraints at low $q^2$ to reduce the number of free parameters.
In this way one separates the problem of the normalisation of the $B^+\to K^+ \mu^+\mu^-$ rate, and the corresponding extraction of short-distance 
Wilson coefficients, from the problem of obtaining a reliable description of the dimuon spectrum. 
While within our approach there is no significant progress on the first problem, there is a tangible advantage on the second one.
The parameterisation of the amplitude we propose is flexible enough to allow the extraction of all the relevant parameters characterising hadronic thresholds in the  dimuon spectrum from data, while retaining significant predictive power in the smooth region within and below the two narrow charmonium resonances. This fact is the key property which allows us to set useful constraints on the $B^+\to K^+ \tau^+\tau^- \to K^+ \mu^+\mu^-$ re-scattering from future precise measurements of  ${\rm d}  \Gamma (B^+\to K^+ \mu^+\mu^-)/ {\rm d} q^2$.

The method we have proposed is  particularly well suited for the LHCb experiment, which has already collected a large sample of $B^+\to K^+ \mu^+\mu^-$ 
events and has an excellent resolution in the dimuon spectrum~\cite{Aaij:2016cbx}.
As we have shown, the data already collected in run II should allow to set a bound on $\cB(B \to K\tau^+\tau^-)$ of $O(10^{-3})$, competitive with current direct bounds (see Table~\ref{tab:sensitivity}). Bounds of $O(10^{-4})$ could be obtained with the LHCb upgrade-II luminosity.

\subsubsection*{Acknowledgments}

This project has received funding from the  European Research Council (ERC) under the European Union's Horizon 2020 research 
and innovation programme  under grant agreement 833280 (FLAY), 
and by the Swiss National Science Foundation (SNF) under contract 200021-159720. 


\end{document}